\documentclass[twocolumn,showpacs,preprintnumbers,amsmath,amssymb,pre,floatfix]{revtex4}

\usepackage{graphicx}
\usepackage{dcolumn}
\usepackage{bm}

\begin{document}

\title{Properties of contact matrices induced by pairwise interactions in
proteins
}

\author{Sanzo Miyazawa}
 \email{sanzo.miyazawa@gmail.com}
 \homepage{http://www.sanzo.org/~miyazawa/}
\affiliation{
Graduate School of Engineering, Gunma University, Kiryu, Gunma 376-8515, Japan
}

\author{Akira R. Kinjo}
 \email{akinjo@protein.osaka-u.ac.jp}
\affiliation{
Institute for Protein Reseach, Osaka University, Suita, Osaka, 565-0871, Japan
}

\date{\today}

\newcommand{\script}[1]{{\mbox{\scriptsize #1}}}
\newcommand{\CITE}[1]{ \cite{#1}}
\newcommand{\VEC}[1]{\mathbf{#1}}
\newcommand{\Eq}[1]{Eq. ({#1})}
\newcommand{\Eqs}[1]{Eqs. ({#1})}
\newcommand{\EqPunc}[1]{{#1}}
\newcommand{\Fig}[1]{Fig. {#1}}
\newcommand{\Figure}[1]{Figure {#1}}
\newcommand{\Figs}[1]{Figures {#1}}
\newcommand{\Table}[1]{Table {#1}}
\newcommand{\Tables}[1]{Tables {#1}}
\newcommand{\Ref}[1]{Ref. {#1}}
\newcommand{\FigureInText}[1]{#1}
\newcommand{\FigureLegends}[1]{}

\bibliographystyle{srt}    

\newcommand{\Begin}[1]{\begin{#1}}	
\newcommand{\End}[1]{\end{#1}}		

\newcommand{\BeginBib}[1]{\begin{thebibliography}{#1}}
\newcommand{\EndBib}{\end{thebibliography}}
\newcommand{\Bibitem}[1]{\bibitem{#1}}	

\newcommand{\FirstAuth}[3]{{#2} {#3} {#1}}	
\newcommand{\FirstAuthLFM}[3]{{#2}. {#3}. {#1}}
\newcommand{\FirstAuthLF}[2]{{#2}. {#1}}
\newcommand{\FirstAuthLFMJ}[4]{{#2}. {#3}. {#4} {#1}}

\newcommand{\Name}[3]{{#2} {#3} {#1}}	
\newcommand{\NameLFM}[3]{{#2}. {#3}. {#1}}
\newcommand{\NameLF}[2]{{#2}. {#1}}
\newcommand{\NameLFMJ}[4]{{#2}. {#3}. {#4} {#1}}

\newcommand{\LastName}[3]{{#2} {#3} {#1},} 
\newcommand{\LastNameLFM}[3]{{#2}. {#3}. {#1},}
\newcommand{\LastNameLF}[2]{{#2}. {#1},}
\newcommand{\LastNameLFMJ}[4]{{#2}. {#3}. {#4} {#1},}

\newcommand{\ALComma}{,}		
\newcommand{\COmma}{\ALComma}		
\newcommand{\ALAnd}{ and }			
\newcommand{\ANd}{\ALAnd}			

\newcommand{\LastAuth}[3]{{#2} {#3} {#1},} 
\newcommand{\LastAuthLFM}[3]{{#2}. {#3}. {#1},}
\newcommand{\LastAuthLF}[2]{{#2}. {#1},}
\newcommand{\LastAuthLFMJ}[4]{{#2}. {#3}. {#4} {#1},}

\newcommand{\AuthListEnd}{,}
\newcommand{\ALEnd}{\AuthListEnd}

\newcommand{\Journal}[6]{{#2} {\textbf{#3}}, {#4} ({#6}).}

\newcommand{\InBook}[7]{p.{#5} in {#2}, ed. {#3} ({#4}, {#7}).}

\newcommand{\Book}[3]{{\textit{#1}} ({#2}, {#3}).}

\newcommand{\AnnRevBiophysBioeng}{Annu. Rev. Biophys. Bioeng.}
\newcommand{\AnnuRevBiophysBioeng}{Annu. Rev. Biophys. Bioeng.}
\newcommand{\BioChem}{Biochemistry}
\newcommand{\Biochem}{Biochemistry}
\newcommand{\Bioinfo}{Bioinformatics}
\newcommand{\BiophysJ}{Biophys. J.}
\newcommand{\Biopoly}{Biopolymers}
\newcommand{\BMCBioinfo}{BMC Bioinformatics}
\newcommand{\BullMathBiol}{Bull. Math. Biol.}
\newcommand{\CABIOS}{CABIOS}
\newcommand{\ChemPhysLett}{Chem. Phys. Lett.}
\newcommand{\EurBiophysJ}{Eur. Biophys. J.}
\newcommand{\FoldDesign}{Folding \& Design}
\newcommand{\FoldDes}{Folding \& Design}
\newcommand{\FoldingDes}{Folding \& Design}
\newcommand{\Genetics}{Genetics}
\newcommand{\IntJPeptProtRes}{Int. J. Peptide Protein Res.}
\newcommand{\IntJQuantChem}{Int. J. Quantum Chem.}
\newcommand{\JApplCryst}{J. of Appl. Crystallogr.}
\newcommand{\JBiochem}{J. Biochem.}
\newcommand{\JBiolChem}{J. Biol. Chem.}
\newcommand{\JChemPhys}{J. Chem. Phys.}
\newcommand{\JCP}{J. Chem. Phys.}
\newcommand{\JCompAidMolDesign}{J. Comp.-Aided Mol. Design}
\newcommand{\JComputChem}{J. Comput. Chem.}
\newcommand{\JMB}{J. Mol. Biol.}
\newcommand{\JMolBiol}{J. Mol. Biol.}
\newcommand{\JME}{J. Mol. Evol.}
\newcommand{\JMolEvol}{J. Mol. Evol.}
\newcommand{\JPeptProtRes}{J. Peptide Protein Res.}
\newcommand{\JPhysChem}{J. Phys. Chem.}
\newcommand{\Macromol}{Macromolecules}
\newcommand{\Macromolecules}{Macromolecules}
\newcommand{\MolBiolEvol}{Mol. Biol. Evol.}
\newcommand{\MolPhylEvol}{Mol. Phyl. Evol.}
\newcommand{\NAR}{Nucl. Acid Res.}
\newcommand{\Nature}{Nature}
\newcommand{\NatureNewBiol}{Nature New Biol.}
\newcommand{\NatureL}{Nature(London)}
\newcommand{\PhysLettB}{{Phys. Lett.} {B}}
\newcommand{\PhysRevLett}{Phys. Rev. Lett.}
\newcommand{\PhysRevD}{{Phys. Rev.} {D}}
\newcommand{\PhysRevE}{{Phys. Rev.} {E}}
\newcommand{\PNAS}{Proc. Natl. Acad. Sci. USA}
\newcommand{\ProtEng}{Protein Eng.}
\newcommand{\ProtSci}{Protein Sci.}
\newcommand{\Proteins}{Proteins}
\newcommand{\QuartRevBiophys}{Quart. Rev. Biophys.}
\newcommand{\Science}{Science}
\newcommand{\SysBiol}{System Biol.}
\newcommand{\ZPhysC}{{Z. Phys.} {C}}

\renewcommand{\VEC}[1]{{\bf #1}}

\begin{abstract}

The properties of contact matrices ($C$ matrices) needed for native proteins 
to be the lowest-energy conformations are considered in relation to
a contact energy matrix ($E$ matrix).
The total conformational energy
is assumed to consist of pairwise interaction energies
between atoms or residues, each of which is
expressed as a product of a conformation-dependent function 
(an element of the $C$ matrix)
and a sequence-dependent energy parameter
(an element of the $E$ matrix).
Such pairwise interactions in proteins force
native $C$ matrices to be in a relationship as if
the interactions are a Go-like potential
[N. Go, Annu. Rev. Biophys. Bioeng. \textbf{12}. 183 (1983)]
for the native $C$ matrix, because
the lowest bound of the total energy function
is equal to the total energy of the native conformation interacting in
a Go-like pairwise potential.
This relationship between $C$ and $E$ matrices 
corresponds to
(a) a parallel relationship
between the eigenvectors of the $C$ and $E$ matrices
and a linear relationship between their eigenvalues,
and (b) a parallel relationship between a contact number vector
and the principal eigenvectors of the $C$ and $E$ matrices,
where the $E$ matrix is expanded
in a series of eigenspaces with an additional constant term.
The additional constant term 
in the spectral expansion of the $E$ matrix 
is indicated by the lowest bound of the total energy function
to correspond to a threshold of contact energy that
approximately separates native contacts from non-native ones.
Inner products between 
the principal eigenvector 
of the $C$ matrix, that of the $E$ matrix, and a contact number vector
have been examined for 182 proteins each of which is a representative 
from each family of the SCOP database
[A. G. Murzin et al., J. Mol. Biol. \textbf{247}, 536 (1995)], 
and
the results indicate the parallel tendencies between those vectors.
A statistical contact potential
[S. Miyazawa and R. L. Jernigan, Proteins \textbf{34}, 49 (1999); 
\textbf{50}, 35 (2003)]
estimated from
protein crystal structures was used to evaluate
pairwise residue-residue interactions in the proteins.
In addition, the spectral representation of 
$C$ and $E$ matrices
reveals that pairwise residue-residue interactions,
which depends only on the types of interacting amino acids but not on other residues
in a protein, are insufficient and other interactions including
residue connectivities and steric hindrance
are needed to make native structures the unique lowest-energy conformations.

\end{abstract}

\maketitle

\section{\label{sec:level1}Introduction}

Predicting a protein three dimensional structure from
its sequence is equivalent to reproducing
a three dimensional structure from one dimensional information
encoded in its sequence. 
From such a viewpoint,
there are many studies that try to reconstruct
three dimensional structures from one dimensional information
such as contact numbers and 
the principal eigenvector of a contact matrix
\CITE{KKVD:02,PBRV:04,KN:05,VWP:06}.
An important question is not only what kind of one dimensional information  
is needed to reconstruct protein structures but also
why such information is critical to reconstruct protein structures.

Let us think about
a distance matrix each element of which is equal to
distance between atoms or residues specified by its column and row.
Information contained in the distance matrix is equivalent with 
the specification of three-dimensional coordinates of each atom or residue,
except that a mirror image of the native structure 
cannot be excluded in distance information.
Reconstructing a distance matrix from one-dimensional vectors requires
in principle the specification of all eigenvectors as well as eigenvalues.
In other words, for an $N \times N$ matrix, $N$ $N$-dimensional
vectors are required.
However, protein's particular characteristics may allow
the reconstruction of a distance matrix with fewer one-dimensional vectors.

A contact matrix whose element is equal to
one for contacting atom or residue pairs or zero for no-contacting atom or residue
pairs 
on the basis of distance between the two atoms/residues,
is a simplification of
a distance matrix with two categories, contact or non-contact, 
but keeps almost all information
needed to reconstruct three-dimensional structures of proteins.
In the case of a residue-residue contact matrix 
consisting of discrete values, one and zero, 
Porto et al.\CITE{PBRV:04}
showed that the contact map of the native structure of
globular proteins can be reconstructed
starting from the sole knowledge of the contact map's principal
eigenvector, and the reconstructed contact map allow in turn
for an accurate reconstruction of the three-dimensional structure.

A vector of contact numbers, which is defined as
the number of atoms or residues in contact with each atom or residue  in a protein,
is another type of one-dimensional vector that is often used
as a one-dimensional representation of protein structures\CITE{NO:80,NO:86,KHN:05},
and may be similar to but not the same as the principal eigenvector
of a contact matrix.
Kabak\c{c}io\v{g}lu et al.\CITE{KKVD:02} suggested that 
the number of feasible protein conformations that
satisfy the constraint of a contact number for each residue is very limited.

A question is why the principal eigenvector of a contact matrix
and a contact number vector contain significant information on
protein structures.
Here, we consider 
what properties of contact matrices are induced 
by pairwise contact interactions 
for native proteins to be the lowest-energy conformations.
For simplicity,
a total conformational energy 
is assumed to consist of pairwise interactions over all atom or residue pairs.
It is further assumed that
the pairwise interaction
can be expressed as 
a product of a conformation-dependent ($C$-dependent) factor and 
a sequence-dependent ($S$-dependent) factor.
The $C$-dependent factor 
represents the degree of contact between atoms/residues
and can be assumed without loss of generality
to take any value between 0 and 1.
The $S$-dependent factor
corresponds to an energy parameter specific to a given pair of atoms or residues.
Here we call a matrix of the $C$-dependent factor
a generalized contact matrix or even simply a contact matrix ($C$ matrix),
and call a matrix of the $S$-dependent factor a generalized contact energy matrix
or even simply a contact energy matrix ($E$ matrix).
A simple linear algebra indicates that
such a total energy function is bounded by the lowest value
corresponding to the total energy for a $C$ matrix
in which all pairs with lower contact energies than
a certain threshold are in contact.
Such a lower bound is achieved\CITE{KM:08} if and only if 
proteins are ideal to have the so-called Go-like potential\CITE{G:83}.
The Go-like potential is defined as the one
in which interaction energies between native contacts are
always lower than those between non-native contacts.
Real pairwise interactions in proteins could not be
the Go-like potential.
In other words, real proteins could not achieve this lowest bound
of a pairwise potential because of atom and residue connectivities and
steric hindrance that are not included in this type of 
total energy function.
How should they approach to the lowest bound as closely as
possible?  
The lowest bound can be approached by
making the singular vectors of the $C$ matrix parallel to
the corresponding singular vectors of the $E$ matrix with the same value of
the singular values. 
Also, in the lowest bound a contact number vector tends to be parallel to
the principal eigenvectors of the $C$ and $E$ matrices.
The most effective way would be to first make the principal
singular vector of the $C$ matrix parallel to that of the $E$ matrix.
A similar strategy was used to recognize protein structures
by three-dimensional threading of protein sequences\CITE{CIWMSDH:04,CWDIH:06}.
Bastolla et al.\CITE{BPRV:05} pointed out that
the principal eigenvector of a contact matrix
must be correlated with that of a contact energy matrix,
if the free energy of a conformation folded into a contact map
is approximated by a pairwise contact potential.
It was shown that 
the correlation coefficients of these two principal eigenvectors
are actually statistically significant in protein folds.
However, unlike their analyses
the lowest bound of the total energy indicates 
the $E$ matrix to be singular-decomposed with a constant term that
corresponds to the threshold energy to separate 
native contacts from non-native ones.
The eigenvectors of $E$ matrix depend 
on the value of the additional constant.

Based on the indication above, we have analyzed   
the relationships between the principal eigenvectors
of the $C$ and $E$ matrices and contact number vector
by examining the inner product of the two vectors.
A statistical contact potential\CITE{MJ:99,MJ:03} estimated from
protein crystal structures is used to evaluate
pairwise residue-residue interactions in proteins.
One hundred and eighty-two representatives of single domain proteins
from each family in the SCOP version 1.69 database\CITE{MBHC:95}
are used to  analyze the relationship between 
the principal eigenvectors of the native $C$ and $E$ matrices
and the contact number vector.
Results show that
the inner product of both the principal eigenvectors has
a maximum at a certain value of the threshold energy 
for contacts, and that there are parallel tendencies between
both the principal eigenvectors and contact number vector.
It is worth noting that
the principal eigenvector of the native $C$-matrix corresponds to
the lower frequency normal modes of the native structure of protein.

In addition, the spectral representation of $C$ and
$E$ matrices
reveals that pairwise residue-residue interactions,
which depend only on the types of interacting amino acids but not on other residues 
in a protein, are insufficient and other interactions including
residue connectivities and steric hindrance
are needed to make native structures unique lowest-energy conformations.

\section{Methods}

\vspace*{2em}
\noindent
\textbf{Basic assumptions and conventions}
\vspace*{1em}

  We first assume that the total conformational energy 
of a protein with conformation $C$ and amino acid sequence $S$
of $N$ units can be approximated as the sum of pairwise interaction energies
between the units.  Here a single unit may consist of an atom or a residue, 
although in most cases we treat a residue as a unit.
We further assume that each pairwise interaction term
can be expressed as a product of a $C$-dependent factor and 
an $S$-dependent factor. 
The $C$-dependent factor represents the degree to which a pair of units are in contact, 
while the $S$-dependent factor represents an interaction energy for a contacting pair of units. 
In other words, the total conformational energy is assumed to be approximated as
\begin{eqnarray}
E^{\script{c}}(C, S) &=& \frac{1}{2} \sum_{i}^N \sum_{j}^N 
	\mathcal{E}_{i j}(S)
	\Delta_{ij}(C)
	\label{eq: total-contact-energy}
	\label{eq: total-contact-energy-A}
	\\
 &=& \frac{1}{2} \sum_{i}^N \sum_{j}^N \delta 
	\mathcal{E}_{i j}(S)
	\Delta_{ij}(C)
	+ \varepsilon_0 N_{\script{c}}(C) \EqPunc{,} 
	\label{eq: total-contact-energy-B}
	\\
\delta 
	\mathcal{E}_{i j}(S)
	&\equiv& 
	\mathcal{E}_{i j}(S)
	- \varepsilon_0 \EqPunc{.}
	\label{eq: def-varepsilon}
\end{eqnarray}
where 
$\mathcal{E}_{i j}(S)$
and $\Delta_{ij}(C)$ are
the $S$-dependent and $C$-dependent factors for the pairwise interaction energy
between the $i$th and $j$th units, respectively. 
$N_{\script{c}}(C)$ is 
the total number of contacts between units and defined as
\begin{eqnarray}
N_{\script{c}}(C) &\equiv& \frac{1}{2} \sum_{i} \sum_{j} \Delta_{ij}(C) = \frac{1}{2} \sum_{i} n_i(C)
	\EqPunc{,}
	\label{eq: def-total-number-of-contacts}
\end{eqnarray}
where the generalized contact number $n_i$,
which is the total number of units contacting with the $i$th unit, is defined as
\begin{eqnarray}
        n_i(C) &=& \sum_j^N \Delta_{ij}(C) \EqPunc{.}	\label{eq: def-contact-number}
\end{eqnarray}
In \Eq{\ref{eq: total-contact-energy-B}},
a constant $\varepsilon_0$ defined by \Eq{\ref{eq: def-varepsilon}}
is introduced to explicitly treat the total number of contacts   
in the evaluation of the total energy.

Each $\Delta_{ij}(C)$ is a function of coordinates of
the $i$th and $j$th units, and
is assumed without loss of generality to take any value between 0 and 1, 
with the diagonal elements always defined to be equal to 0. 
The $S$-dependent term 
$\mathcal{E}_{i j}(S)$
can include not only
two-body interactions but multi-body effects such as a mean-field,
that is, it can not only depend on the type of a unit pair but
on the entire protein sequence.
We call the matrix $\Delta(C) \equiv (\Delta_{ij}(C))$ as a generalized
contact matrix or $C$-matrix for short. 
Similarly, we call the matrix 
$(\mathcal{E}_{i j}(S))$ 
as a generalized contact energy matrix or $E$-matrix for short.
Each element of the energy function of \Eq{\ref{eq: total-contact-energy-A}}
can represent either attractive or repulsive interactions
but not both.  
In the next sections, we consider the mathematical lower limits
of the total contact energy, ignoring atomic details of proteins 
such as atom and residue connectivities and steric hindrance.
The volume exclusions between atoms are assumed
to be satisfied and are not included in the total energy function.
To minimally reflect the effects of steric hindrance,
the total number of contacts $N_{\script{c}}$ is explicitly
treated in the evaluation of the total energy, 
\Eq{\ref{eq: total-contact-energy-B}}, by introducing
a constant $\varepsilon_0$. 
The expression for \Eq{\ref{eq: total-contact-energy}}
can be regarded as a special case of \Eq{\ref{eq: total-contact-energy-B}}
in which $\varepsilon_0$ is equal to zero.

\vspace*{2em}
\noindent
\textbf{Lower bounds of the total contact energy} 
\vspace*{1em}

Let us consider lower bounds of the total contact energy 
represented by \Eq{\ref{eq: total-contact-energy}} 
under a condition that
each element of $C$-matrix can independently take any value 
within $0 \leq \Delta_{ij} \leq 1$
irrespective of whether or not they can be reached in real protein conformations;
in other words, atom and residue connectivities and steric hindrance
are completely ignored. 

If one regards 
$\delta \mathcal{E}_{i j}$
and $\Delta_{ij}$ as the elements of the vectors 
$\delta \vec{\mathcal{E}}(S)$ 
and $\vec{\Delta}(C)$ 
in $N^2$-dimensional Euclidean space, 
it will be obvious that
the first term of \Eq{\ref{eq: total-contact-energy-B}}
can be bounded by a product of the norms of those two vectors:
\begin{eqnarray}
E^{\script{c}}(C, S)
	&\geq& 
	\!
	\min_{\varepsilon_0} \, [ \,
	- \, \frac{1}{2}
	\| \delta \vec{\mathcal{E}}(S)  \|
	\| \vec{\Delta}(C) \|
	+ \varepsilon_0 N_{\script{c}}(C) 
	\, ]
	\EqPunc{,}
	\label{eq: unreachable-lowest-bound-A}
	\label{eq: unreachable-lowest-bound-AA}
\end{eqnarray}
where $\| \ldots \|$ means a Euclidian norm.
Obviously the equality of \Eq{\ref{eq: unreachable-lowest-bound-A}} 
is achieved if and only if
those vectors are anti-parallel to each other:
\begin{eqnarray}
\delta \mathcal{E}_{i j}(S)
	&=& \varepsilon \Delta_{ij}(C) \EqPunc{,}
	\label{eq: go-type-1}
	\label{eq: go-type-1A}
\end{eqnarray}
where $\varepsilon $ is a negative constant.

In addition, there is a simple mathematical limit for
the total energy of \Eq{\ref{eq: total-contact-energy}} for which
the $C$ matrix is equal to 
$H_0(- \delta \mathcal{E}_{i j})$:
\begin{eqnarray}
\lefteqn{
E^{\script{c}}(C, S)
}
	\nonumber \\
	&\geq& 
	\!
	\frac{1}{2} \sum_{i} \sum_{j} 
	\delta \mathcal{E}_{i j}(S)
	\Delta_{ij}(C_{\script{min}})
	+ \varepsilon_0(C_{\script{min}}) N_{\script{c}}(C_{\script{min}})
	\label{eq: unreachable-lowest-bound-B}
	\\
	&\geq& 
	\!
	\frac{1}{2} \sum_{i} \sum_{j} 
	\mathcal{E}_{i j}(S)
	H_0(- \mathcal{E}_{i j}(S) ) \EqPunc{,}
	\label{eq: unreachable-lowest-bound}
	\label{eq: unreachable-lowest-bound-C}
	\\
\lefteqn{
\Delta_{ij}(C_{\script{min}}) =	
	H_0(- \delta \mathcal{E}_{i j}(S)) \EqPunc{,}
}
	\label{eq: go-type-2}
\end{eqnarray}
where
$H_0(x)$ is the Heaviside step function that takes 1 for $x > 0$ and
0 for otherwise. 
$C_{\script{min}}$ is
the lowest-energy conformation with
a constraint on the total contact number $N_{\script{c}}$, 
although it is not necessarily reached due to
atom and residue connectivities, and steric hindrance.
If each $\Delta_{ij}$ is allowed to take either 0 or 1 only,
and also each $\delta \varepsilon_{ij}$ takes 
either one of two real values only
to be able to satisfy \Eq{\ref{eq: go-type-1}},
both the lower bounds of
\Eqs{\ref{eq: unreachable-lowest-bound-A}}
and \ref{eq: unreachable-lowest-bound-B}
are equal to each other.
Otherwise, 
the lower bound of \Eq{\ref{eq: unreachable-lowest-bound-A}} is 
further bounded by the lower bound of \Eq{\ref{eq: unreachable-lowest-bound-B}}, 
or
the equality in \Eq{\ref{eq: unreachable-lowest-bound-A}} cannot be achieved 
with $0 \leq \Delta_{ij} \leq 1$,
but \Eq{\ref{eq: unreachable-lowest-bound-B}} is always satisfied.
If the total number of contacts $N_{\script{c}}$ is constrained to be equal to 
$N_{\script{c}}(C_{\script{min}})$, then
$\varepsilon_0$ must be properly chosen as a non-positive value
so that \Eq{\ref{eq: def-total-number-of-contacts}} is satisfied
with $C = C_{\script{min}}$.
Otherwise, $\varepsilon_0$ should be taken to be 
equal to 0 to obtain the lower bound of
\Eq{\ref{eq: unreachable-lowest-bound}}.
\Eq{\ref{eq: unreachable-lowest-bound}}
describes the lowest bound 
without any constraint on the number of contacts
and corresponds to the energy of the conformation $C_{\script{min}}$
for the case of $\varepsilon_0 = 0$.

The potentials that satisfy
\Eq{\ref{eq: go-type-1}} or \ref{eq: go-type-2} are just 
Go-like potentials\CITE{G:83}, 
in which interactions between native contact pairs are always more attractive
than those between non-native pairs.
Let us call proteins with a Go-like potential as
ideal proteins.
There are multiple levels of nativelikeliness in the Go-like potential.
The most nativelike potential of the present Go-like potentials is
the one in which all interactions between native contacts are attractive
and other interactions are all repulsive.
In other words, 
$\mathcal{E}_{i j}$ 
is negative for native contacts and positive for non-native contacts.
In such a Go-like potential, the native conformation can attain the 
lowest bound of \Eq{\ref{eq: unreachable-lowest-bound-C}},
which is equivalent to \Eq{\ref{eq: unreachable-lowest-bound-B}}
with $\varepsilon_0 = 0$.
A less nativelike potential is the one in which
interactions between non-native contact pairs can be attractive but
always less attractive than those between native contact pairs.
An ideal protein with such a potential
can attain \Eq{\ref{eq: unreachable-lowest-bound-B}}
with a proper value of $\varepsilon_0$, which
is the threshold energy for native and non-native contacts.
For real protein, we should define $\varepsilon_0$ as a threshold
of contact energy under which unit pairs tend to be in contact
in native conformations.

In ideal proteins, the lowest-energy conformation must be
the one for which the contact potential
looks like a Go-like potential,
and inversely
the potential must be 
a Go-like potential for the lowest-energy conformation.
In real proteins,
it would be impossible that
contact potentials for native structures are exactly like
a Go-like potential of
\Eq{\ref{eq: go-type-1}} or \Eq{\ref{eq: go-type-2}},
even though the contact potential being considered here
may be the effective one that includes not only actual pairwise interactions 
but also the effects of higher order interactions near native structures.
In other words, the lowest bound of \Eq{\ref{eq: unreachable-lowest-bound-B}}
could not be achieved for real pairwise potentials,
because of atom and residue connectivities and steric hindrance.
However, it is desirable to reduce frustrations
among interactions so that an effective pairwise
potential in native structures must approach
the Go-like potential. 
Then, a question is 
how native contact energies approach the mathematical lowest limit.
In the following, we will give tips as to how the $C$-matrix should
be designed to decrease the total energy towards the theoretical lowest limit.

It should be noted here that the lowest-energy conformation, the $C$ matrix, is considered
for a given potential, the $E$ matrix, but not its inverse problem, which is
to consider an optimum potential or an optimum sequence
for a given conformation --- that is, an optimum $E$ matrix for a given $C$ matrix.
In the inverse problem, the total partition function varies depending on
each sequence, and it must be taken into account to evaluate the stability of
the given $C$ matrix in relative to the other conformations \CITE{SVMB:96,DK:96,MoS:96,MJ:99a}. 
The $Z$ score of the energy gap between the given $C$ matrix and other compact conformations
may be used to evaluate the optimality of each sequence \CITE{MAS:96,BPRV:05}.

\vspace*{2em}

\noindent
\textbf{Spectral relationship between $C$  and $E$ matrices} 
\vspace*{1em}

We apply singular value decomposition to both the $C$ matrix 
(generalized contact matrix) and $E$ matrix 
(generalized contact energy matrix).
The $C$ matrix is decomposed as
\begin{eqnarray}
\Delta_{ij}(C) &=& \sum_{\mu} | \lambda_{\mu}(C) | L_{i \mu}(C) R_{j \mu}(C)  \EqPunc{,}
			\label{eq: eigen-equation-for-contact-matrix} 
		\\
| \lambda_{1}(C) | &\geq& \ldots \geq  | \lambda_{N}(C)| \geq 0  \EqPunc{,}
\end{eqnarray}
where
$\lambda_{\mu}(C)$ is the eigenvalue of $\Delta(C)$, and its absolute value
$|\lambda_{\mu}(C)|$ 
is the $\mu$th non-negative singular value of $\Delta(C)$ arranged in decreasing order,
and $\VEC{L}_{\mu}(C) \equiv \ ^t(L_{1 \mu}, \ldots, L_{N \mu})$ and 
$\VEC{R}_{\mu}(C) \equiv \ ^t(R_{1 \mu}, \ldots, R_{N \mu})$ are
the corresponding left and right singular vectors;
both $L \equiv (\VEC{L}_{1}, \ldots, \VEC{L}_{N})$ and $R \equiv (\VEC{R}_{1}, \ldots, \VEC{R}_{N})$ are orthonormal matrices.
Note that the singular values for a symmetric matrix such as a contact matrix 
are equal to the absolute value of its eigenvalue. 
We choose 
the eigenvector corresponding to the eigenvalue $\lambda_{\mu}(C)$ 
as a right singular vector $\VEC{R}_{\mu}(C)$
and if $\lambda_{\mu}(C) \geq 0$, $\VEC{L}_{\mu}(C) \equiv \VEC{R}_{\mu}(C)$ and
otherwise $\VEC{L}_{\mu}(C) \equiv - \; \VEC{R}_{\mu}(C)$.

Likewise, the $E$ matrix, 
$(\mathcal{E}_{i j}(S))$, 
is decomposed as
\begin{eqnarray}
\mathcal{E}_{i j}(S)
	&=& \sum_{\nu} | \varepsilon_{\nu} | U_{i\nu}(S) V_{j \nu}(S) + \varepsilon_0  \EqPunc{,}
	\label{eq: eigen-equation-for-E-matrix}
		\\
|\varepsilon_{1}| &\geq& \ldots \geq | \varepsilon_{N}| \geq 0  \EqPunc{,}
\end{eqnarray}
where  the absolute value of the eigenvalue, $|\varepsilon_{\nu}(S)|$, 
$\VEC{U}_{\nu}(S) \equiv \ ^t(U_{1\nu}, \ldots, U_{N\nu})$, and 
$\VEC{V}_{\nu}(S) \equiv \ ^t(V_{1\nu}, \ldots, V_{N\nu})$ are 
the $\nu$th singular value, left singular vector, and right singular vector of 
the matrix 
$(\delta \mathcal{E}_{i j}(S))$, 
respectively.
We choose 
the eigenvector corresponding to the eigenvalue $\varepsilon_{\nu}(C)$ 
as a right singular vector $\VEC{V}_{\nu}(C)$
and if $\varepsilon_{\nu}(C) \geq 0$, $\VEC{U}_{\nu}(C) \equiv \VEC{V}_{\nu}(C)$ and
otherwise $\VEC{U}_{\nu}(C) \equiv - \; \VEC{V}_{\nu}(C)$.

 We then substitute 
\Eqs{\ref{eq: eigen-equation-for-contact-matrix}} and \ref{eq: eigen-equation-for-E-matrix} into the definition of the total energy, \Eq{\ref{eq: total-contact-energy}},
and obtain
\begin{eqnarray}
E^{\script{c}}(C, S)
	&=& \frac{1}{2} \sum_{\mu} \sum_{\nu} | \lambda_{\mu}(C)||\varepsilon_{\nu}(S)| \omega_{\mu \nu}(C, S) 
	\nonumber \\ 
	& & + \varepsilon_0 N_{\script{c}}(C)  \EqPunc{,}
	\label{eq: total-contact-energy-2}
\end{eqnarray}
where
\begin{eqnarray}
\omega_{\mu \nu}(C,S)  &\equiv& \sum_i L_{i \mu}(C) U_{i \nu}(S) \sum_j R_{j \mu}(C) V_{j \nu}(S)
	\nonumber 
	\\
	&=& ^t\VEC{L}_{\mu}(C) \VEC{U}_{\nu}(S)  ^t\VEC{R}_{\mu}(C) \VEC{V}_{\nu}(S)  \EqPunc{.}
\end{eqnarray}
Because the first term in \Eq{\ref{eq: total-contact-energy-2}} is
simply the trace of the product of two matrices, 
$\mbox{tr} \; (\delta \mathcal{E} ^t\Delta)$,
Neumann's trace theorem \cite{HJ:85} leads to the following inequality:
\begin{eqnarray}
\lefteqn{
E^{\script{c}}(C, S)
}
	\nonumber \\
         &\geq&
	\!
	\min_{\varepsilon_0} \, [ \,
	 - \, \frac{1}{2} 
                \sum_{\{\xi | \lambda_{\xi} \varepsilon_{\xi} \neq 0 \}}
                | \lambda_{\xi}(C)
                \varepsilon_{\xi}(S) |
		+ \varepsilon_0 N_{\script{c}}(C)
	\, ]
	\EqPunc{.}
        \label{eq: lowest-bound-of-energy-eigenvector-A}
        \label{eq: lowest-bound-of-energy-eigenvector-A-2}
        \label{eq: lowest-bound-of-energy-eigenvector-B}
        \label{eq: lowest-bound-of-energy-eigenvector-C}
	\label{eq: lowest-bound-of-energy-eigenvector-D}
\end{eqnarray}
The equality in \Eq{\ref{eq: lowest-bound-of-energy-eigenvector-C}} is achieved
if and only if
\begin{eqnarray}
	\omega_{\mu \nu} &=& 
	- \delta_{\mu \nu}
	\hspace*{1em} 
	\mbox{ for } \{ \mu | \lambda_{\mu} \varepsilon_{\mu} \neq 0 \}  \EqPunc{,}
	\label{eq: best-optimum-principal-eigenvector}
	\label{eq: best-optimum-eigenvectors}
	\label{eq: best-optimum-singular-vectors}
\end{eqnarray}
that is, all the corresponding left and right singular vectors of the $C$- and $E$-matrices
are exactly parallel or anti-parallel to each other.
Then, regarding the singular values as the elements of a vector --- i.e.,
$\vec{\lambda}(C) \equiv {^t( \lambda_1, \ldots, \lambda_N )}$ 
and $\vec{\varepsilon}(S) \equiv {^t( \varepsilon_1, \ldots, \varepsilon_N )}$ ---
the sum of the products of the eigenvalues 
of the $E$  and $C$ matrices
in \Eq{\ref{eq: lowest-bound-of-energy-eigenvector-D}}
can be bounded 
by the product of the norms of those two vectors, which is
equal to the product of the norms of the vectors consisting of
$E$  or $C$ matrix elements.
As a result, 
we obtain the lower bound corresponding to \Eq{\ref{eq: unreachable-lowest-bound-A}} 
already derived in the previous section:
\begin{eqnarray}
E^{\script{c}}(C, S)
	&\geq& 
	\min_{\varepsilon_0} \, [ \,
	- \, \frac{1}{2} 
	\| \vec{\lambda}(C) \|_{\{ \xi | \lambda_{\xi} \varepsilon_{\xi} \neq 0 \}}
	\| \vec{\varepsilon}(S) \|_{\{ \xi | \lambda_{\xi} \varepsilon_{\xi} \neq 0 \}}
\nonumber \\
		 & & + \varepsilon_0 N_{\script{c}}(C)
	\; ]
	\label{eq: lowest-bound-of-energy-eigenvector-E}
	\\
	&=& 
	\min_{\varepsilon_0} \, [ \,
	 - \, \frac{1}{2}
	\| \delta \vec{\mathcal{E}}(S) \|_{\{ \xi | \lambda_{\xi} \varepsilon_{\xi} \neq 0 \}}
	\| \vec{\Delta}(C) \|_{\{ \xi | \lambda_{\xi} \varepsilon_{\xi} \neq 0 \}}
\nonumber \\
	& & + \varepsilon_0 N_{\script{c}}(C)
	\; ]
	\EqPunc{,}
	\label{eq: unreachable-lowest-bound-AAA}
\end{eqnarray}
where $\| \cdots \|_{\{ \xi | \lambda_{\xi} \varepsilon_{\xi} \neq 0 \}}$
means the norm in the subspace of $\lambda_{\xi} \varepsilon_{\xi} \neq 0$.
The equality of	
\Eq{\ref{eq: lowest-bound-of-energy-eigenvector-E}}
is achieved
if and only if
the values of the eigenvalues of the $C$ matrix are proportional to
those of the $E$ matrix:
\begin{eqnarray}
	\varepsilon_{\xi}(S) &=& \varepsilon \lambda_{\xi}(C)
	\hspace*{1em}
	\mbox{ for } \{ \xi | \lambda_{\xi} \varepsilon_{\xi} \neq 0 \}  \EqPunc{.}
	\label{eq: go-type-1B}
\end{eqnarray}
Note that $\varepsilon $ is a negative constant
due to \Eq{\ref{eq: best-optimum-singular-vectors}}.
This condition with \Eq{\ref{eq: best-optimum-singular-vectors}} 
corresponds to \Eq{\ref{eq: go-type-1A}}, but
the spectral representation of $C$ and $E$ matrices
reveals that 
the relation of \Eq{\ref{eq: go-type-1B}} is required only
for the eigenspaces of $\lambda_{\xi} \varepsilon_{\xi} \neq 0$.
 
\vspace*{3em}
\noindent
\textbf{Is a pairwise residue-residue potential sufficient 
to make native structures unique lowest-energy conformations ?}
\vspace*{1em}

If there exists $\xi$ such that $\varepsilon_{\xi} = 0$,
and the $C$-matrices for two conformations $C$ and $C'$ satisfy
$(^tU (\Delta(C) - \Delta(C')) V)_{\xi\xi} = 0$
for $\{\xi | \varepsilon_{\xi} \neq 0 \}$ and $N_{\script{c}}(C) = N_{\script{c}}(C')$,
those two conformations have the same conformational energy,
because the total contact energy can be represented as
\begin{eqnarray}
E^{\script{c}}(C,S) &=& \! \frac{1}{2} \sum_{\nu} | \varepsilon_{\nu} | (^tU \Delta(C) V )_{\nu\nu}
	+ \varepsilon_0 N_{\script{c}}(C)  \EqPunc{.}
\end{eqnarray}
If the contact interactions are genuine two-body between residues, 
$\mathcal{E}_{i j}(S)$ and $\delta \mathcal{E}_{i j}(S)$ 
will depend only on the residue type of the $i$th and $j$th
units and 
therefore $\mbox{rank} (\delta \mathcal{E}_{i j})$ will be
less than or equal to the number of amino acid types in
a protein; therefore,
$\mbox{rank} (\delta \mathcal{E}_{i j}) \leq 20$.
Thus, in the case of genuine two-body interactions between residues,
there must exist $\xi$ such that $\varepsilon_{\xi} = 0$ for any
chain longer than 20 residues --- that is, multiple $C$ matrices with
the same energy.
In other words, interactions other than pairwise interactions
are needed to make native structures unique lowest-energy conformations.
A certain success\CITE{MJ:05} of genuine two-body statistical potentials
in identifying native structures as unique lowest-energy conformations
indicates that most of the eigenspaces of $\varepsilon_{\xi} = 0$,
especially in orientation-dependent potentials,  
may be significantly reduced or even
disallowed for short proteins by atom and residue connectivities and steric hindrance. 
It may be worthy of note that
the number of possible $C$-matrices is of the order of $2^{N(N-1)/2}$ but
the conformational entropy of self-avoiding chains is proportional
to at most $N$, where $N$ is the chain length;
that is, vast conformational space becomes disallowed by
chain connectivity and steric hindrance.
However, it would be not surprising even if
a two-body contact potential is insufficient to make 
all the native structures be unique lowest-energy conformations,
especially for long amino acid sequences.
Actually it was reported\cite{MS:96,TE:00,TSLE:00} that it is impossible to
optimize a pairwise potential to identify all native structures.
Multi-body interactions\CITE{MS:97}
may be required as a mean-field or even explicitly
together with the two-body interactions,
as well as other interactions 
such as secondary structure potentials\CITE{CFT:06}.

\vspace*{3em}
\noindent
\textbf{Relationship between a contact number vector $\VEC{n}$ and eigenvectors of the $C$ matrix}
\vspace*{1em}

\Eq{\ref{eq: lowest-bound-of-energy-eigenvector-D}} indicates that
the larger the principal eigenvalue is,  
the lower is the lower bound of the total contact energy. 
The eigenvalue $\lambda_\mu$ satisfies
\begin{eqnarray}
\lambda_{\mu}(C) &=& 
	 \frac{^t\VEC{R}_{\mu}(C) \VEC{n}(C)}{^t\VEC{R}_{\mu}(C) \VEC{1}}
		\\
		&=& \langle n_{\bullet}^2 \rangle^{1/2} \ ^t\VEC{R}_{\mu} \VEC{n} \| \VEC{1} \| / ( ^t\VEC{R}_{\mu} \VEC{1} \| \VEC{n} \| )
		\EqPunc{,}
		\label{eq: eigenvalue-vs-contact-number}
\end{eqnarray}
where $^t\VEC{R}_{\mu} \VEC{n} / \| \VEC{n} \|$ is 
the cosine of the angle between the contact number vector $\VEC{n}$
and eigenvector $\VEC{R}_{\mu}$, and 
$^t\VEC{R}_{\mu} \VEC{1} / \| \VEC{1} \|$ 
is the one between the eigenvector $\VEC{R}_{\mu}$ and the vector $\VEC{1}$ whose elements are all equal to 1. 
Here
$\langle n_{\bullet}^2 \rangle$ represents the second moment of contact numbers over all units.
We can say that the eigenvalue $\lambda_{\mu}$ is equal to 
the weighted average of contact number $n_i$ 
with each component of the eigenvector, $R_{i\mu}$, 
and also that it is roughly proportional to the square root of the second moment of
contact numbers.
The principal eigenvalue has a value within the range of
$ 2 N_{\script{c}} / N \leq \lambda_{1} \leq \max_i n_i$\CITE{B:98}.
The larger 
the ratio 
$^t\VEC{R}_{\mu} \VEC{n} \| \VEC{1} \| / ( ^t\VEC{R}_{\mu} \VEC{1} \| \VEC{n} \| )$
is, the larger the eigenvalue $\lambda_{\mu}$ becomes. 
It has been reported that the contact number vector is
highly correlated with the principal eigenvector of 
the $C$ matrix\CITE{PBRV:04,KN:05}.

\vspace*{2em}
\noindent
\textbf{Relationship between a contact number vector $\VEC{n}$ and 
eigenvectors of the $E$-matrix}
\vspace*{1em}

A contact number vector is a $C$ matrix summed 
over a row or column. 
Thus, to obtain a relationship between the contact number vector $\VEC{n}$
and eigenvectors of the $E$ matrix, an averaging of the $E$ matrix over a row or column
is needed.

We approximate the total contact energy as follows by
replacing 
$\delta \mathcal{E}_{i j}$ 
by its average over the index $j$, 
$\delta \mathcal{E}_{i \bullet}$, 
and then obtain an approximate expression for the lower bound of the total contact energy:
\begin{eqnarray}
\lefteqn{
E^{\script{c}}(C, S) 
}
\nonumber \\
&\approx& 
	\frac{1}{2} \sum_i \sum_j [ \frac{1}{N} \sum_k 
\delta \mathcal{E}_{i k}(S) 
	]
	\Delta_{ij}(C) 
	+ \varepsilon_0 N_{\script{c}}(C)
	\label{eq: approx_total_contact_energy-0}
	\\
	&=& \frac{1}{2} \;
	^t\delta \vec{\mathcal{E}}_{\bullet}(S) 
	\VEC{n}(C)
	+ \varepsilon_0 N_{\script{c}}(C)
		\label{eq: approx_total_contact_energy}
	\\
	&\geq& 
	- \; \frac{1}{2} 
	\| \delta \vec{\mathcal{E}}_{\bullet}(S) \| 
	\| \VEC{n}(C)\|
	+ \varepsilon_0 N_{\script{c}}(C)
		\EqPunc{,}
		\label{eq: mean-contact-energy-vs-contact-number-1}
\end{eqnarray}
where the mean contact energy vector $\delta \vec{\mathcal{E}}_{\bullet}$ 
is defined as
$\delta \vec{\mathcal{E}}_{\bullet}(S) 
\equiv ^t(\ldots, \frac{1}{N} \sum_k 
\delta \mathcal{E}_{i k}(S), \ldots)$.
The equality in \Eq{\ref{eq: mean-contact-energy-vs-contact-number-1}}
holds if and only if the two vectors
$\delta \vec{\mathcal{E}}_{\bullet}$
and $\VEC{n}$ are anti-parallel:
\begin{eqnarray}
	\frac{\delta \vec{\mathcal{E}}_{\bullet}(S)}{\| \delta \vec{\mathcal{E}}_{\bullet}(S) \|}
	&=& - \; \frac{\VEC{n}(C)}{\| \VEC{n}(C) \|}
		\EqPunc{.}
		\label{eq: mean-contact-energy-vs-contact-number}
\end{eqnarray}
\Eq{\ref{eq: mean-contact-energy-vs-contact-number}} above is equivalent to 
the following relation between the contact number vector and 
the eigenvector of the $E$ matrix:
\begin{eqnarray}
\frac{^t\VEC{V}_{\nu} \VEC{n} \| \VEC{1} \| }{^t\VEC{V}_{\nu} \VEC{1} \| \VEC{n} \| }
	&=& 
	\frac{- \varepsilon_{\nu}}{ (\sum_{\nu} (\varepsilon_{\nu} \ ^t\VEC{V}_{\nu} \VEC{1} / \| \VEC{1} \| ) )^2 )^{1/2}}
		\EqPunc{.}
		\label{eq: eigenvector-vs-contact-number}
\end{eqnarray}
If the $E$ matrix can be well approximated by the principal eigenvector term only,
then this condition 
leads to the parallel orientation between
$\VEC{n}$ and the principal eigenvector of $E$-matrix, that is,
$^t\VEC{V}_{1} \VEC{n} / \| \VEC{n} \|  \simeq 1$.

If the conformation for the lower bound of the total energy is 
also the lower-bound conformation even for this averaging over the $E$ matrix,  
\Eq{\ref{eq: mean-contact-energy-vs-contact-number}} or
\ref{eq: eigenvector-vs-contact-number}
above together with \Eq{\ref{eq: best-optimum-eigenvectors}} and
\ref{eq: eigenvalue-vs-contact-number},
$\VEC{n} = \sum_{\mu} \lambda_{\mu} \VEC{R}_{\mu}(^t\VEC{U}_{\mu} \VEC{1})$
and
$\delta \vec{\mathcal{E}} 
= \sum_{\nu} \varepsilon_{\nu} \VEC{V}_{\nu}(^t\VEC{V}_{\nu} \VEC{1})$, leads to
\Eq{\ref{eq: go-type-1B}}
between the eigenvalues of the $C$ and $E$ matrices as follows:
\begin{eqnarray}
\lefteqn{
\lambda_{\xi} (C)
}
	\nonumber
	\\
	&\approx&
	\frac{- \; (\sum_{\xi} (\lambda_{\xi} \ ^t\VEC{R}_{\xi} \VEC{1} / \| \VEC{1} \| )^2 )^{1/2} \varepsilon_{\xi}}{ (\sum_{\xi} (\varepsilon_{\xi} \ ^t\VEC{V}_{\xi} \VEC{1} / \| \VEC{1} \| )^2 )^{1/2}}
	\mbox{ if } {R}_{\xi} = \pm \VEC{V}_{\xi}
	\label{eq: lambda_for_global_minimum_0}
	\\
	&=& \frac{\varepsilon_{\xi}}{\varepsilon}   \hspace*{2ex} \hspace*{2ex} \mbox{ with a negative constant, } \varepsilon < 0  
	\EqPunc{,}
	\label{eq: lambda_for_global_minimum}
\end{eqnarray}
where $\varepsilon$ is a constant taking any negative value.

\section{Data analyses}

\Eq{\ref{eq: lowest-bound-of-energy-eigenvector-D}}
indicates that 
with an optimum value for $\varepsilon_0$
the spectral relationship of
\Eq{\ref{eq: best-optimum-singular-vectors}}
between $E$ and $C$ matrices
tends to be satisfied in the lowest-energy conformations.
Here we will examine it by crudely evaluating
pairwise interactions with a contact potential between
amino acids, which was estimated as a statistical
potential from contact frequencies between amino acids 
observed in protein crystal structures.

\vspace*{2em}
\noindent
\textbf{Pairwise contact potential used} 

A contact potential used is
a statistical estimate\CITE{MJ:03} of contact energies 
with a correction\CITE{MJ:99} for the Bethe approximation\CITE{MJ:85,MJ:96}.
The contact energy between amino acids of type $a$ and $b$ was estimated as
\begin{eqnarray}
e_{ab} &=& e_{rr} + \alpha' [ \Delta e^{\script{Bethe}}_{ar} + \Delta e^{\script{Bethe}}_{rb} 
	+ \frac{\beta'}{\alpha'} \delta e^{\script{Bethe}}_{ab}]
	\EqPunc{.}
	\label{eq: contact_potential_B}
\end{eqnarray}
$e_{rr}$ is part of contact energies irrespective of residue types and is called a collapse energy, which 
is essential for a protein to fold by cancelling out the large
conformational entropy of extended conformations but
cannot be estimated explicitly from contact frequencies 
between amino acids in protein structures.
$\Delta e^{\script{Bethe}}_{ar}$ and $\delta e^{\script{Bethe}}_{ab}$ are
the values of $\Delta e_{ar}$ and $\delta e_{ab}$
evaluated by the Bethe approximation 
from the observed numbers of contacts between amino acids.
$\Delta e_{ar} + e_{rr}$ is a partition energy or hydrophobic energy for
a residue of type $a$.  $\delta e_{ab}$ is an intrinsic contact energy
for a contact between residues of type $a$ and $b$;
refer to\CITE{MJ:99} for those exact definitions. 
The proportional constants for correction
were estimated as $\beta'/ \alpha' = 2.2$ and 
$\alpha' \leq 1$\CITE{MJ:99}.
Here energy is measured in $kT$ units; $k$ is the Boltzmann constant
and $T$ is the temperature.
With the spectral expansion of the second term of \Eq{\ref{eq: contact_potential_B}},
the contact energies can be represented by 
\begin{eqnarray}
e_{ab} &=& e_{rr} + \alpha' [ \sum_{\nu} e_{\nu} Q_{a\nu} Q_{b\nu} + e_0 ] 
	\EqPunc{,}
	\label{eq: eigen_system_for_contact_energy}
\end{eqnarray}
where $e_{\nu}$ and $\VEC{Q}_{\nu}$ are eigenvalues and eigenvectors
for the second term of \Eq{\ref{eq: contact_potential_B}} 
with a constant $e_0$.
Li et al.\CITE{LTW:97} showed that 
the contact potential\CITE{MJ:85,MJ:96} corresponding to 
$\beta' / \alpha' = 1$  between residues
can be well approximated by 
the principal eigenvector term together with a constant term.

Then, the following relationship is derived for the eigenvalues and
eigenvectors between the $E$ matrix $(\mathcal{E}_{i j})$ 
and the contact energy matrix $(e_{ab})$:
\begin{eqnarray}
\varepsilon_{0} &=& e_{rr} +  \alpha' e_0
	\EqPunc{,}
	\label{eq: def_of_varepsilon_0}
	\\
\varepsilon_{\nu} &\approx& \alpha' e_{\nu} \sum_i Q^2_{a_i \nu} = \alpha' e_{\nu} \langle Q^2_{a_i \nu} \rangle  N 
	\EqPunc{,}
	\label{eq: def_of_varepsilon_1}
	\\
V_{i \nu} &\approx& Q_{a_i \nu} / (\sum_i Q^2_{a_i \nu})^{1/2}
	\EqPunc{,}
\end{eqnarray}
where $a_i$ is the amino acid type of the $i$th residue, and $N$ is the protein length.
It should be noted here that the eigenvectors $\VEC{V}_{\nu}$ do not depend on the value of $\alpha'$.

The $C$ matrix $\Delta(C)$ is defined
in such a way that non-diagonal elements 
take a value 1 for residues that are completely in contact,
a value 0 for residues that are too far from each other, and values between 1 and 0
for residues whose distance is intermediate between those two extremes.
Contacts between neighboring residues are completely ignored, that is
$\Delta_{i j} = 0$ for $|i -j| \leq 1$.
The geometric center of side chain heavy atoms or the $C_{\alpha}$ atom for glycine 
is used to represent each residue.
Previously, this function was defined as a step function for simplicity.
Here, it is defined as a switching function as follows;
in the equation below to define residue contacts, $\VEC{r}_i$ means 
the position vector of a geometric center of side chain heavy atoms
or the $C^{\alpha}$ atom for glycine:
\begin{eqnarray}
\Delta(\VEC{r}_i, \VEC{r}_j) &\equiv&
	S_w(|\VEC{r}_i - \VEC{r}_j|, d^c_1, d^c_2)
	\EqPunc{,}
				 \\
S_w(x, a, b) &\equiv& 
			\left\{ \begin{array}{l}
	1	\hspace*{2em}	\mbox{ for $x \le a$} 	\\
	\! [ (b^2 - x^2)^2 / (b^2 - a^2)^3 ]	\\
	\hspace*{1em}	 \times \ [ 3(b^2 - a^2) - 2 (b^2 - x^2) ]
			\\
	\ 	\hspace*{2em}   \mbox{ for $a < x < b$}  \EqPunc{,} \\
	0	\hspace*{2em}	\mbox{ for $b \le x$}
			\end{array}
			\right.			
			\label{eq: def_Sw}
\end{eqnarray}
where $S_w$ is a switching function that sharply changes its value from 
1 to 0 between the lower distance $d^c_1$ and the upper distance $d^c_2$. 
Those critical distances $d^c_1$ and $d^c_2$ are taken here as
$6.65$  \AA \ and $7.35$  \AA, respectively.
\vspace*{2ex}

\vspace*{2ex}
\noindent
{\bf Protein structures analyzed}
\vspace*{1ex}

Proteins each of which is a single-domain protein 
representing a different
family of 
protein folds 
were collected.
In the case of multi-domain proteins in which
contacts between domains are significantly less that
those within domains,
a contact matrix could be approximated by a direct sum of
subspaces corresponding to each domain.  This characteristic 
of multi-domain proteins has been used 
for domain decomposition\CITE{HS:94} and
for identification of side-chain clusters in a protein\CITE{KV:99,KV:00}.
Thus, only single-domain proteins are used here.
Release 1.69 of the SCOP database
\CITE{MBHC:95}
was used for the classification of protein folds.
We have assumed that 
proteins whose
domain specifications
in the SCOP database consist of protein ID only, 
are single-domain proteins.
Representatives of
families
are the first entries in
the protein lists 
for 
each family
in the SCOP; 
if these first proteins in the lists are not appropriate
(see below) to use, 
for the present purpose, 
then the second ones are chosen.
These 
species are all those belonging to
the protein classes 1 -- 4 --- that is, classes of 
all $\alpha$, all $\beta$,
$\alpha/\beta$, and $\alpha+\beta$
proteins.
Classes of multi-domain, membrane
and cell surface proteins, small proteins, peptides and designed proteins
are not used.
Proteins whose structures\CITE{BWFGBWSB:00} were determined by NMR or
having stated resolutions worse than 2.0 \AA \ are removed 
to assure that the quality of proteins used is high.
Also, proteins whose coordinate sets consist either of only $C^\alpha$ atoms,
or include many unknown residues, or lack many atoms or residues,
are removed. In addition, proteins shorter than 50 residues are also removed.
As a result, 
the set of family representatives includes
182 protein domains.

\section{Results}

The spectral relationship between the $C$
and $E$ matrices
is analyzed for
single domain proteins that are representatives
from each family of classes 1 -- 4 in the SCOP
database of version 1.69.
The statistical potential used is crude, so that
the following analyses are limited only to relationships between
the principal eigenvectors of the $C$ and 
$E$ matrices and contact number vector.
It should be noted here that a crude evaluation of 
the pairwise interactions may make their relationships unclear.

\FigureInText{

\begin{figure}[ht]
\centerline{
\includegraphics*[width=80mm,angle=0]{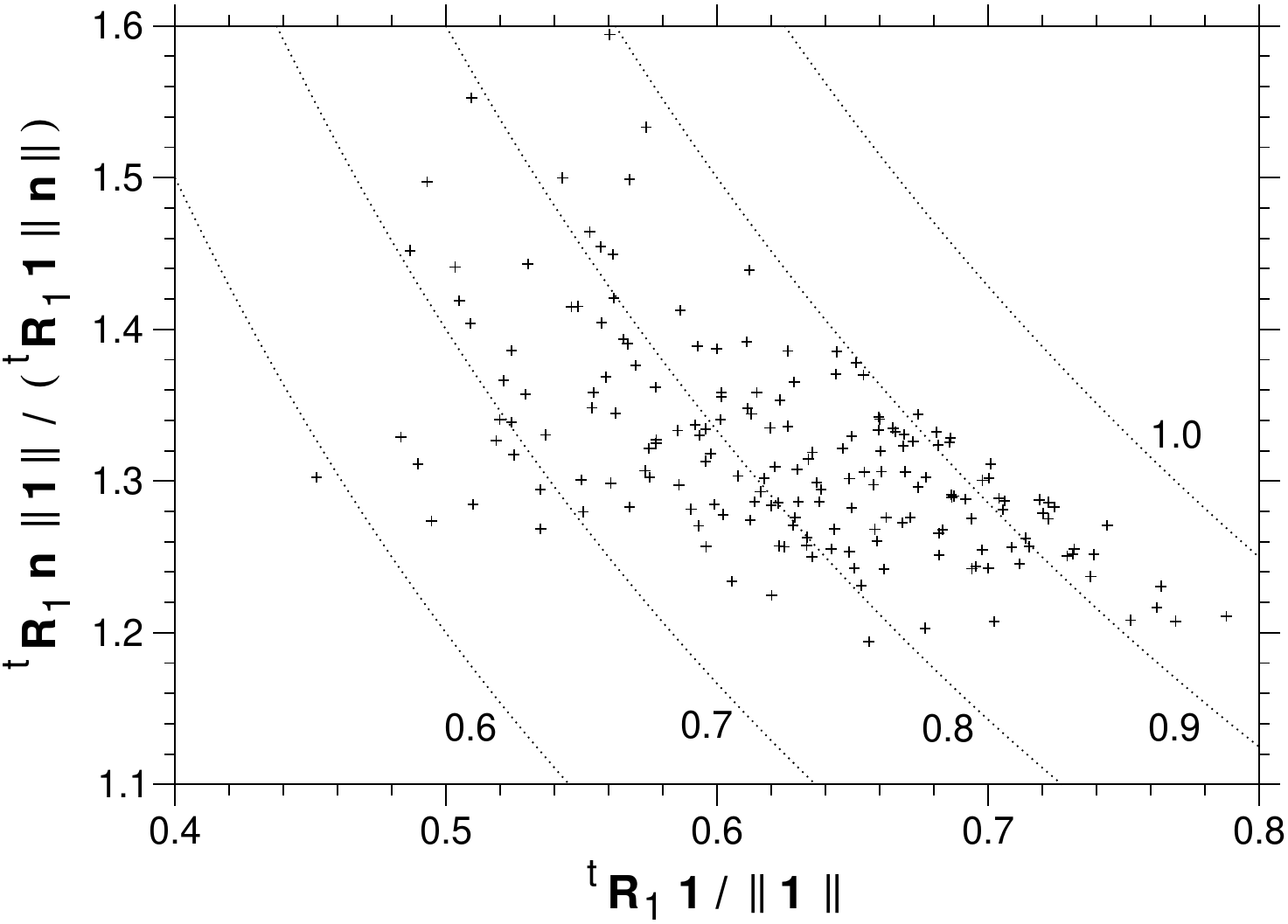}
}
\caption{\label{fig: cos_cpe_cn_vs_sqrt_n_ave_cpei}
The ratio of $^t\VEC{R}_{1} \VEC{n} / \parallel \VEC{n} \parallel$ to $^t\VEC{R}_{1} \VEC{1} / \parallel \VEC{1} \parallel$
is shown for each of 182 proteins,
which are representatives of
single domain proteins from each family of classes 1 -- 4 in the SCOP version 1.69.
$\VEC{R}_{1}$ and $\VEC{n}$ are
the principal eigenvector and contact number vector
of the native $C$ matrix, respectively.
The dotted lines indicate the iso-value lines for 
$^t\VEC{R}_{1} \VEC{n} / \parallel \VEC{n} \parallel$,
whose values are shown in the figure.
}
\end{figure}

} 

\Eq{\ref{eq: eigenvalue-vs-contact-number}}
indicates that the eigenvalues of the $C$ matrix are proportional to
the square root of the second moment
of contact numbers.
The proportional coefficient for the principal eigenvalue of the $C$ matrix --- that is, 
$^t\VEC{R}_1 \VEC{n} \parallel \VEC{1} \parallel / (^t\VEC{R}_1 \VEC{1}  \parallel \VEC{n} \parallel)$ ---
is plotted for each protein in \Fig{\ref{fig: cos_cpe_cn_vs_sqrt_n_ave_cpei}}.
The dotted lines are iso-cosine lines for the angle between
the principal eigenvector of the $C$ matrix and contact number vector,
whose values are written in the figure.
The ratios are scattered between 1.2 and 1.6, 
although the value of the ratio depends on the value of
the abscissa, 
$^t\VEC{R}_{1} \VEC{1} / \parallel \VEC{1} \parallel$.
The cosine of the angle is upper bounded by the value of 1, and therefore
the value of the ratio of the cosines 
becomes correlated with the value of the denominator of the ratio --- i.e., 
$^t\VEC{R}_{1} \VEC{1} / \parallel \VEC{1} \parallel$.
The important fact is that the ratio takes values larger than 1, making
the principal eigenvalue larger.
Here, it should be noted that
the lower bound of the conformational energy linearly depends on
the principal eigenvalue of the $C$ matrix; 
see \Eq{\ref{eq: lowest-bound-of-energy-eigenvector-D}}. 
Thus, the larger the principal eigenvalue is,
the lower the conformational energy becomes.
In practice, this condition seems to yield 
a high correlation between the principal eigenvector and the 
contact number vector;
most of the values of the 
$^t\VEC{R}_{1} \VEC{n} / \parallel \VEC{n} \parallel$,
are greater than 0.7.

\FigureInText{

\begin{figure}[ht]
\centerline{
\includegraphics*[width=80mm,angle=0]{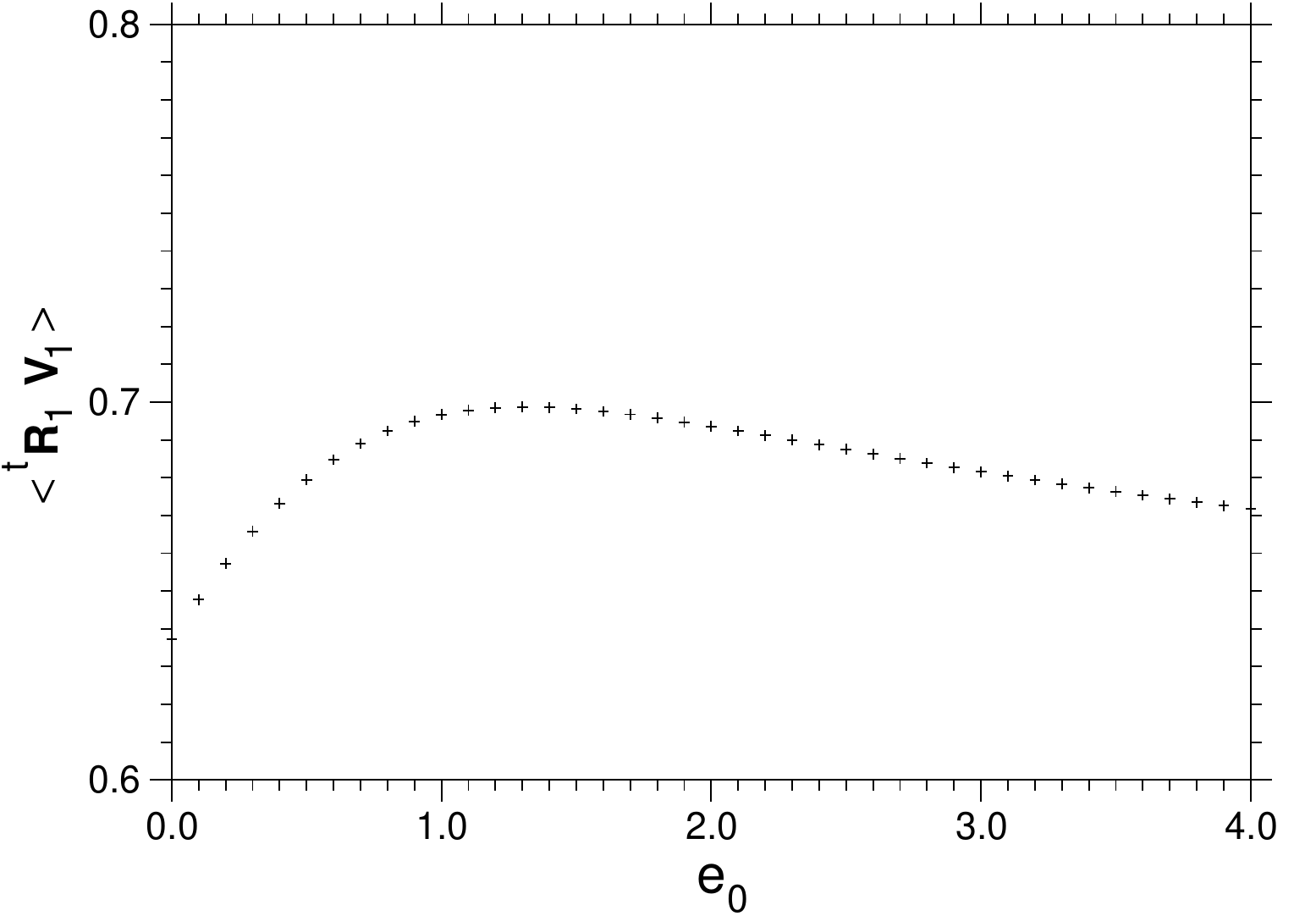}
}
\caption{\label{fig: mean_cos_cpe_epe_vs_e0_scc7.0_22}
The mean of 
$^t\VEC{R}_{1} \VEC{V}_{1}$
over 182 proteins is plotted with plus marks against
$e_0$.
These proteins are representatives of
single domain proteins from each family of classes 1 -- 4 in the SCOP version 1.69.
$\VEC{R}_{1}$ 
is the principal eigenvector of the native $C$ matrix.
$\VEC{V}_{1}$ is the principal eigenvector for the $E$ matrix with the value of $e_0$
specified on the abscissa.
}
\end{figure}

} 

Now let us think about the relationship between the $C$ matrix
and pairwise interactions.
Pairwise interactions between residues are evaluated by using
a statistical estimate\CITE{MJ:03} of contact energies 
with a correction\CITE{MJ:99} for the Bethe approximation.
\Figure{\ref{fig: mean_cos_cpe_epe_vs_e0_scc7.0_22}}
shows the average of 
$^t\VEC{R}_1 \VEC{V}_1$
over all the proteins for each value of $e_0$.  
The average 
$\langle ^t\VEC{R}_1 \VEC{V}_1 \rangle$
takes the maximum value $0.699$ at $e_0 = 1.3$, although 
its decrements according to the increase of $e_0$ are not large.
In the following, $e_0 = 1.3$ is used to calculate 
the eigenvectors of the $E$ matrices.

\FigureInText{

\begin{figure}[ht]
\centerline{
\includegraphics*[height=80mm,angle=0]{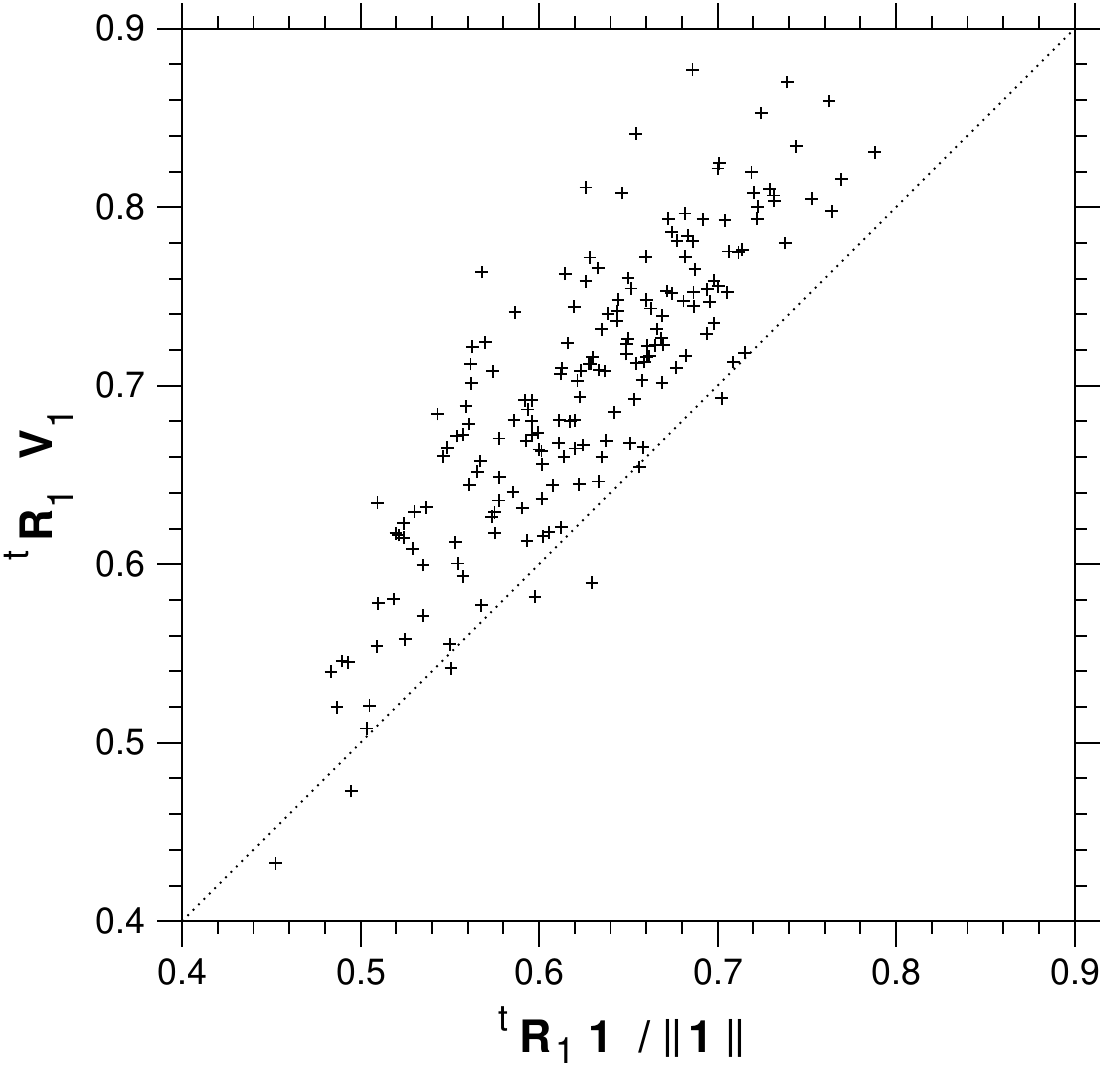}
}
\caption{\label{fig: cos_cpe_epe_vs_sqrt_n_ave_cpei_scc70_22}
The value of 
$^t\VEC{R}_{1} \VEC{V}_{1}$
is plotted against $^t\VEC{R}_{1} \VEC{1} / \| \VEC{1} \|$ 
for each of 182 proteins, 
which are representatives of
single domain proteins from each family of classes 1 -- 4 in the SCOP version 1.69.
$\VEC{R}_{1}$ 
is the principal eigenvector of the native $C$-matrix.
$\VEC{V}_{1}$ is the principal eigenvector for $E$-matrix with $e_0 = 1.3$.
The dotted line shows the line of equal values between the ordinate and abscissa.
}
\end{figure}

} 

The value of 
$^t\VEC{R}_1 \VEC{V}_1$ 
for each protein
is plotted against the value of 
$^t\VEC{R}_1 \VEC{1} / \| \VEC{1} \|$ 
in \Fig{\ref{fig: cos_cpe_epe_vs_sqrt_n_ave_cpei_scc70_22}}.
The value of 
$^t\VEC{R}_1 \VEC{V}_1$ 
is larger 
for most of the proteins than that of 
$^t\VEC{R}_1 \VEC{1} / \| \VEC{1} \|$.
If the direction of 
$\VEC{R}_1$ 
is randomly distributed in the domain of 
$R_{i1} > 0$, 
the probability that 
$^t\VEC{R}_1 \VEC{V}_1$
is larger than 
$^t\VEC{R}_1 \VEC{1} / \| \VEC{1} \|$ 
must be smaller than $0.5$.
Then, in such a random distribution,
the probability to observe 
\Fig{\ref{fig: cos_cpe_epe_vs_sqrt_n_ave_cpei_scc70_22}},
in which 175 of 182 proteins fall into the region of 
$^t\VEC{R}_1 \VEC{V}_1 > \ ^t\VEC{R}_1 \VEC{1}  / \| \VEC{1} \|$,
must be smaller than $_{182}C_{175} (0.5)^{175} = \exp (-91.6)$.
Also t-tests are performed for the correlation coefficients between 
$\VEC{R}_1$ 
and $\VEC{V}_1$
in all proteins. 
The geometric mean of probabilities for a significance 
over 182 proteins examined here is equal to 
$\exp(-18.4)$.
Thus, it is statistically significant that
the direction of the vector 
$\VEC{R}_1$ 
is closer to $\VEC{V}_1$ rather than $\VEC{1}$ whose elements 
do not depend on residues in proteins, This fact indicates
that 
a parallel orientation between 
the principal eigenvectors of the $C$ and $E$ matrices
is favored.

\FigureInText{

\begin{figure}[ht]
\centerline{
\includegraphics*[width=80mm,angle=0]{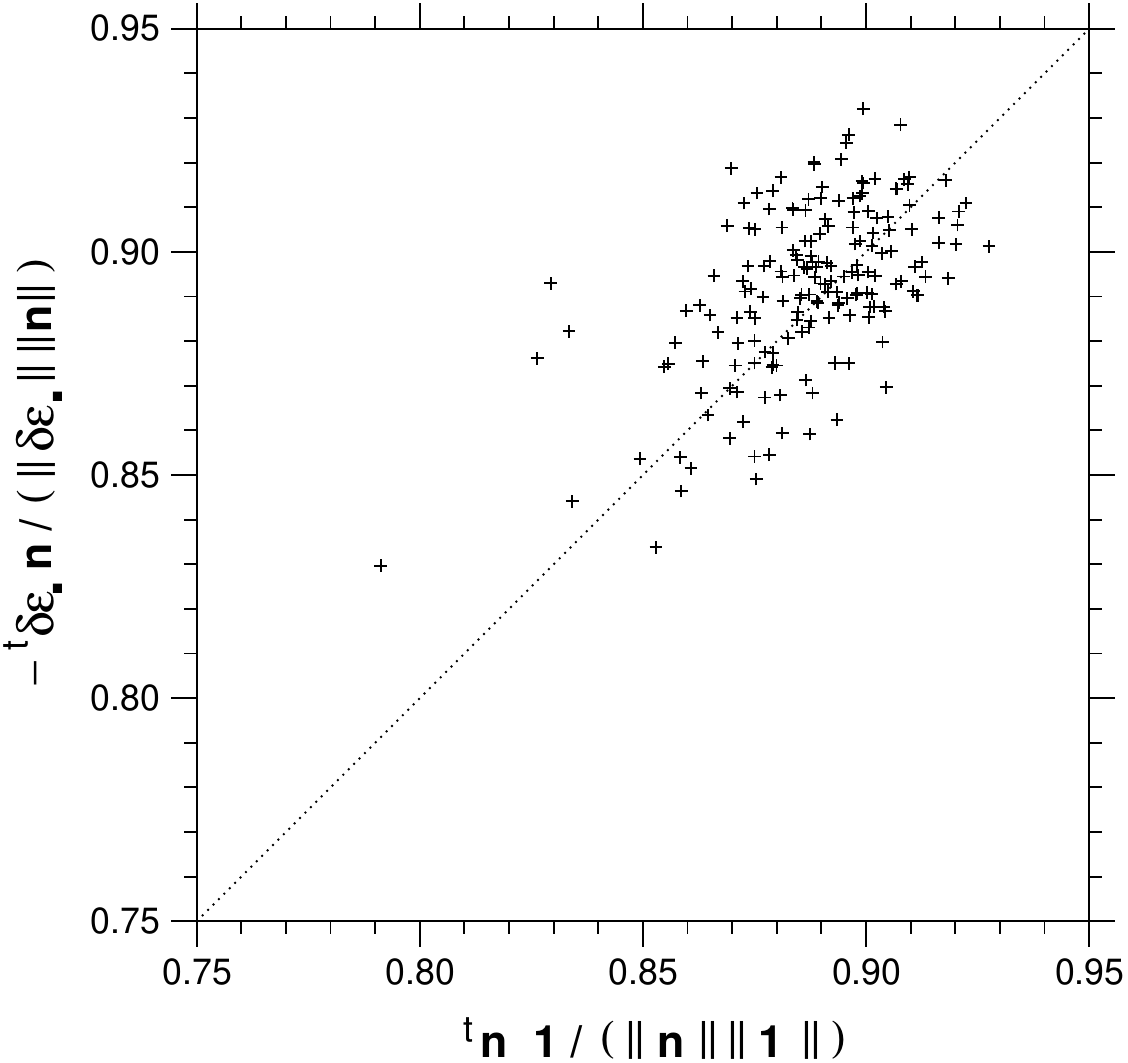}
}
\caption{\label{fig: cos_ave_e_cn_vs_cos_ave_e_1}
The value of 
$- ^t\delta \vec{\mathcal{E}}_{\bullet} \VEC{n} / (\parallel \delta \vec{\mathcal{E}}_{\bullet} \parallel \parallel \VEC{n} \parallel)$
is plotted against 
$^t\VEC{n} \VEC{1} / ( \parallel \VEC{n} \parallel \parallel \VEC{1} \parallel)$ 
for each of 182 proteins,
which are representatives of
single domain proteins from each family of classes 1 -- 4 in the SCOP version 1.69.
$e_0 = 1.3$ is used for the $E$-matrix.
The dotted line shows the line of equal values between the ordinate and abscissa.
}
\end{figure}

} 

\Eq{\ref{eq: mean-contact-energy-vs-contact-number}}
indicates that 
the mean contact energy vector 
$\delta \vec{\mathcal{E}}_{\bullet} ( \equiv 
^t(\ldots, \frac{1}{N} \sum_k 
 \delta \mathcal{E}_{ik }(S), \ldots))$
being antiparallel to 
the contact number vector
is favorable to decrease the conformational energy.
\Figure{\ref{fig: cos_ave_e_cn_vs_cos_ave_e_1}} 
does not show a strong but statistically significant tendency that the value of
$- ^t\delta \vec{\mathcal{E}}_{\bullet} \VEC{n} / ( \parallel \vec{\mathcal{E}}_{\bullet} \parallel \parallel \VEC{n} \parallel) $
tends to be larger than $ ^t \VEC{n} \VEC{1} / (\parallel \VEC{n} \parallel \parallel \VEC{1} \parallel)$;
in t-tests for correlation coefficients between 
$\delta \vec{\mathcal{E}}_{\bullet}$ and $\VEC{n}$,
the geometric mean of probabilities for a significance 
over 182 proteins is equal to 
$\exp(-27.9)$.
If the $E$ matrix can be approximated by the principal eigenvector term,
this fact indicates that the contact number vector
tends to be parallel to the principal eigenvector 
of the $E$ matrix. Actually this is the case for the present estimate
of the contact energies;
the figure of 
$^t\VEC{V}_1 \VEC{n} / \parallel \VEC{n} \parallel$ versus $^t\VEC{n} \VEC{1} / (\parallel \VEC{n} \parallel \parallel \VEC{1} \parallel)$
is not shown.
In t-tests for correlation coefficients between 
$\VEC{V}_1$ and $\VEC{n}$,
the geometric mean of probabilities for a significance 
is equal to 
$\exp(-28.8)$.

\FigureInText{

\begin{figure}[ht]
\centerline{
\includegraphics*[width=80mm,angle=0]{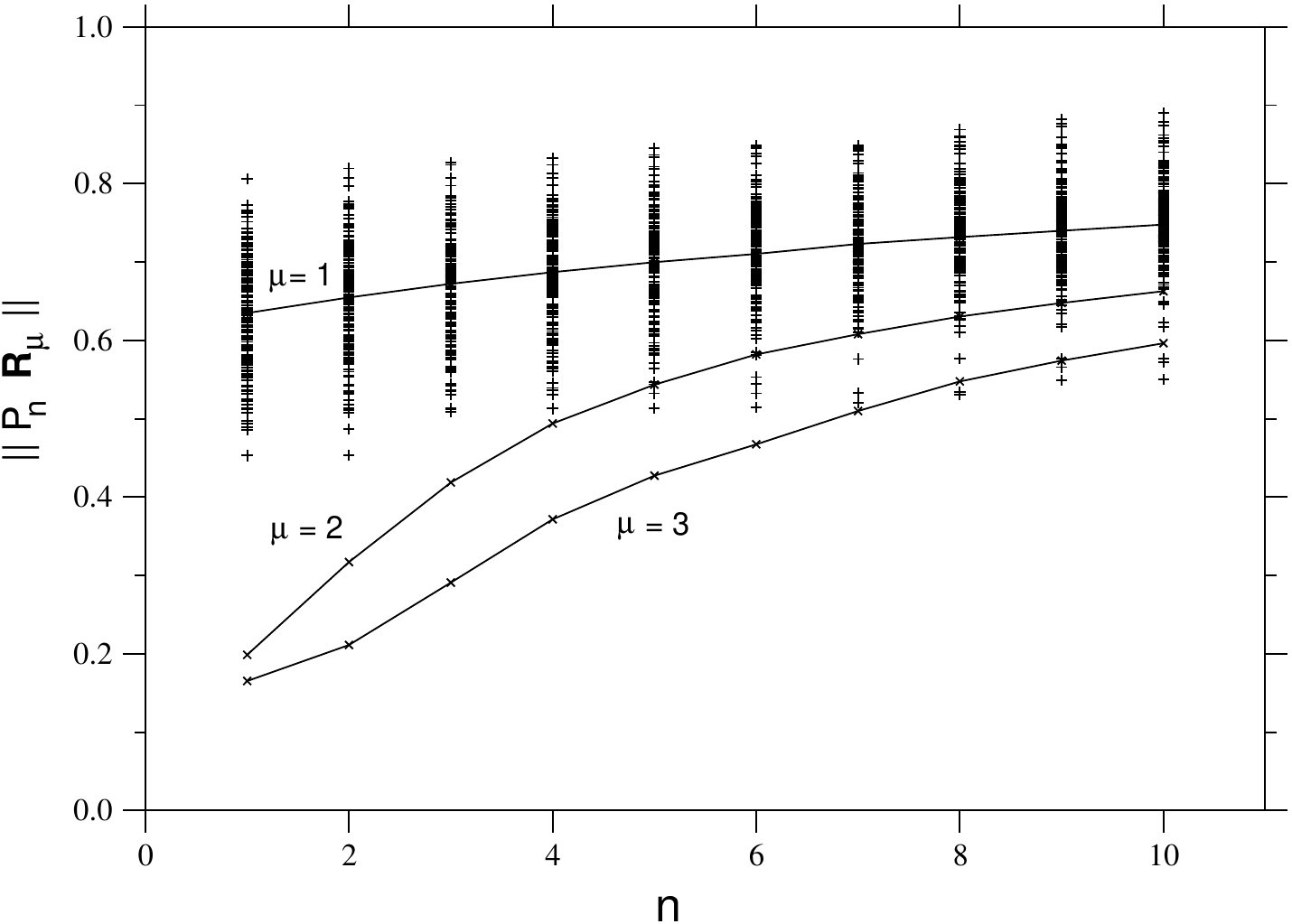}
}
\caption{\label{fig: proj_of_cpe_on_normal_modes}
The norms of the $C$ matrix eigenvectors, $\VEC{R}_{\mu}$,  projected on 
the subspace consisting of the $n$ lowest-frequency normal modes of a Kirchihoff matrix
corresponding to the $C$-matrix, are plotted against $n$; 
the primary eigenspace, indicated by $n=1$, of the Kirchihoff matrix is 
the one consisting of all eigenvalues equal to zero.
$P_n$ means a projection operator on the $n$ lowest-frequency normal modes of 
the Kirchhoff matrix.
Plus marks indicate the norm of the principal eigenvector of
the $C$-matrix of each of 182 proteins
projected on each subspace consisting of the $n$ lowest-frequency normal modes indicated on the abscissa.
The solid curves with cross marks indicate those norms averaged over all the proteins;
their curves from the left to the right 
show those values for the first, the second, and the third principal eigenvectors of the $C$ matrix,
respectively.
}
\end{figure}

} 

Here, we have shown that the principal eigenvector among other
eigenvectors of the $C$ matrix
seems to be a main contributor to minimize the conformation energy.
It is important to take notice that 
the principal eigenvector of the $C$ matrix corresponds to 
the lower-frequency normal modes of protein motion.
Let us think about a Kirchhoff matrix that is defined as
\begin{eqnarray}
K_{ij} &\equiv& n_i \delta_{ij} - \Delta_{ij}  \EqPunc{,}
\end{eqnarray}
where $\delta_{ij}$ is a Kronecker's delta.
The eigenvalue of the Kirchhoff matrix
is equal to the square of normal mode angular frequency in a system
in which $i$th and $j$th units are connected to each other 
by a spring with a spring constant equal to $\Delta_{ij}$.
If the contact number $n_i$ is equal to a constant $n_c$ irrespective of
unit $i$, then the eigenvalue of the Kirchhoff matrix is
equal to $n_c - \lambda_{\mu}$. In other words,
in this case the principal eigenvector of the $C$ matrix 
corresponding to the largest eigenvalue 
is equal to
the eigenvector of the Kirchhoff matrix corresponding to the smallest
eigenvalue --- that is, the lowest-frequency normal mode
corresponding to a motion that leads to a large conformational change\CITE{T:96}.
In actual proteins, the contact number $n_i$ depends on the unit $i$, and then
the correspondence between the eigenvectors of the $C$ matrix and
the Kirchhoff matrix would become vague, but it will be expected that
the principal eigenvector of the $C$ matrix belongs
to a subspace consisting of lower-frequency normal modes.

In \Fig{\ref{fig: proj_of_cpe_on_normal_modes}},
plus marks indicate the norm of the principal eigenvector of the $C$ matrix of each of 182 proteins
projected on each subspace consisting of the $n$ lowest-frequency normal modes indicated on the abscissa.
In most of the proteins, the principal eigenvector of the $C$ matrix corresponds to
the lower-frequency normal modes of the Kirchhoff matrix.
The solid curves with cross marks indicate those norms averaged over all
proteins;
their curves from the left to the right 
show those values for the first, second, and third principal eigenvectors of the $C$ matrix,
respectively.
The solid curve for the principal eigenvector shows that 
about 70\% of the principal eigen vector of the $C$ matrix 
can be explained by only ten lowest-frequency normal modes.
Thus, the principal eigenvector of the $C$ matrix 
is not only an important contributor to minimize conformation energy,
but also corresponds to the lower-frequency normal modes of protein motion.

\vspace*{2em}
\noindent
\section{Discussion}

The lower bounds of the total contact energy lead to 
a relationship between $E$ and $C$ matrices such that
the contact potential looks like a Go-like potential.
Such a relationship may be realized only for ideal proteins, but
in real proteins, atom- and residue-connectivities and
steric hindrance not included in the contact energy
can significantly reduce conformational space;
the number of possible $C$-matrices is of the order of $2^{N(N-1)/2}$ but
the conformational entropy of self-avoiding chains is proportional
to at most $N$, where $N$ is the chain length.
As a result, 
\Eq{\ref{eq: best-optimum-eigenvectors}} 
is expected to be
approximately satisfied only for some singular spaces, probably
for singular values taking relatively large values,
but at least for the principal singular space.
It was confirmed in the representative proteins
that the inner products of 
the principal eigenvectors of $E$ and $C$ matrices 
are significantly biased toward the value 1
at a certain value of the threshold energy $\varepsilon_0$ for contacts,
where their average over all proteins has a maximum; 
see \Fig{\ref{fig: cos_cpe_epe_vs_sqrt_n_ave_cpei_scc70_22}}. 
Parallel relationships were also indicated and confirmed between
the principal eigenvector $\VEC{R}_{1}$
and the contact number vector $\VEC{n}$
of the $C$ matrix 
and between 
the mean contact energy vector $\delta \vec{\mathcal{E}}_{\bullet}$ and
the contact number vector $\VEC{n}$; 
see \Figs{\ref{fig: cos_cpe_cn_vs_sqrt_n_ave_cpei}}
and \ref{fig: cos_ave_e_cn_vs_cos_ave_e_1}.
In these analyses, a statistical potential was used to
evaluate the contact energies between residues, and
the coarse grain of the evaluations 
limits the present analysis to a relationship between
the principal eigenvectors of the $E$ and $C$ matrices,
and also can make the relationship between these matrices vague. 
However, the results clarify 
the significance of the principal eigenvectors
of the $E$ and $C$ matrices and contact number vector 
in protein structures.
Here, it may be worthy of note that
the principal eigenvector of the $C$ matrix corresponds to
the lower-frequency normal modes of protein structures.

The condition for the lowest bound of energy, 
\Eq{\ref{eq: go-type-2}}, 
indicates that 
$\varepsilon_0$ in real proteins 
corresponds to a threshold of contact energy
for a unit pair to tend to be in contact in the native structures. 
In principle, such a threshold for contact energy 
depends on the size of the protein and protein architecture;
it should be noted that
many types of interactions in real proteins are missed in
representing interactions by contact potentials.
The estimate of $e_0$ shown in 
\Fig{\ref{fig: mean_cos_cpe_epe_vs_e0_scc7.0_22}}
is an estimate only for the present specific type of a contact potential.
The important things are that the total contact energy is bounded
by 
\Eq{\ref{eq: unreachable-lowest-bound-B}} 
with a constant term,
and that spectral relationships 
of \Eqs{\ref{eq: best-optimum-singular-vectors}}
and \ref{eq: go-type-1B}
between 
$E$ and $C$ matrices are expected for the conformations 
of the lower bounds if the $E$ matrix is decomposed
with a constant term as shown in 
\Eq{\ref{eq: eigen-equation-for-E-matrix}}.

Besides that,
the spectral representation of $C$ and
$E$ matrices
reveals that pairwise residue-residue interactions,
which depends only on the types of interacting amino acids but not on other residues
in a protein, are insufficient and other interactions including
residue connectivities and steric hindrance
are needed to make native structures unique lowest-energy conformations.

\end{document}